\documentclass[a4paper,12pt]{article}

\usepackage{amsmath,amssymb,mathtools}
\usepackage{color}
\usepackage{graphicx}
\usepackage{subfigure}
\usepackage{cite}
\usepackage{hyperref}
\usepackage{multirow,makecell}
\usepackage{textcomp}
\usepackage{wasysym}
\usepackage{setspace}
\usepackage{verbatim}
\usepackage{float}
\usepackage{ulem}
\usepackage[utf8]{inputenc}
\graphicspath{{fig-magnetic/}}

\usepackage[text={17.2cm,25.2cm},centering]{geometry}

\numberwithin{equation}{section}

\begin{document}
	\title{\textbf{Holographic spin alignment of $J/\psi$ meson in magnetized plasma}}
	\author{Yan-Qing Zhao $^{1}$\footnote{zhaoyanqing@mails.ccnu.edu.cn}, Xin-Li Sheng $^{2}$\footnote{sheng@fi.infn.it}, Si-Wen Li $^{3}$\footnote{siwenli@dlmu.edu.cn}, Defu Hou $^{1}$\footnote{houdf@mail.ccnu.edu.cn}}
	\date{}
	
	\maketitle
	
	\vspace{-10mm}
	
	\begin{center}
		{\it
			$^{1}$ Institute of Particle Physics and Key Laboratory of Quark and Lepton Physics (MOS), \\Central China Normal University, Wuhan 430079, China\\ \vspace{1mm}
		    $^{2}$ INFN Sezione di Firenze, Via G. Sansone 1, I-50019, Sesto F.no (Firenze), Italy\\ \vspace{1mm}
                $^{3}$ Department of Physics, College of Science, Dalian Maritime University, Dalian 116026, China}
            \vspace{10mm}
	\end{center}

	\begin{abstract}
We study the mass spectra and spin alignment of vector meson $J/\psi$ in a thermal magnetized background using a generalized theoretical framework based on gauge/gravity duality. Utilizing a soft wall model for the QGP background and a massive vector field for the $J/\psi$ meson, we delve into the meson's spectral function and spin parameters $(\lambda_{\theta},\, \lambda_\varphi,\,\lambda_{\theta\varphi})$ for different cases, assessing their response to variations in magnetic field strength, momentum, and temperature. We initially examine scenarios where a meson's momentum aligns parallel to the magnetic field in helicity frame. Our results reveal a magnetic field-induced positive $\lambda_\theta^\text{H}$ for low meson momentum, transitioning to negative with increased momentum. As a comparison, we also study the case of momentum perpendicular to the magnetic field and find the direction of magnetic field does not affect the qualitative behavior for the $eB$-dependence of $\lambda_\theta^\text{H}$. Moreover, we apply our model to real heavy-ion collisions for three different spin quantization directions. Further comparisons with experimental data show qualitative agreement for spin parameters $\lambda_{\theta}$ and $\lambda_\varphi$ in the helicity and Collins-Soper frames.
\end{abstract}
	
	\baselineskip 18pt
	\thispagestyle{empty}
	\newpage
	
	\tableofcontents

 \clearpage

\section{Introduction}\label{sec:00_intro}
Relativistic heavy-ion collisions generate a new phase of hot and dense matter called the quark-gluon plasma (QGP). In non-central collisions, there exists a sufficiently strong magnetic field in the center region of the QGP predominantly induced by the positively charged spectators. The magnetic field in Au-Au collisions at $\sqrt{s_{\text{NN}}}=200$ GeV at the Relativistic Heavy Ion Collider (RHIC) in BNL can reach a peak value of approximately 0.1 $\text{GeV}^2$, while that in Pb-Pb collisions at the Large Hadron Collider (LHC) can reach 1 $\text{GeV}^2$ \cite{Skokov:2009qp, Kharzeev:2007jp}. As the spectators swiftly exit the center region post-collision, the magnetic field quickly wanes after the collision. The electrical conductivity of the medium \cite{McLerran:2013hla, Gursoy:2014aka, Tuchin:2014iua, Li:2016tel, Chen:2021nxs, Yan:2021zjc, Wang:2021oqq} can potentially influence the magnetic fields in such a way as to give rise to predicted, yet not experimentally confirmed, phenomena such as the chiral magnetic effect \cite{Kharzeev:2007jp, Fukushima:2008xe, Son:2009tf} and chiral magnetic wave \cite{Kharzeev:2010gd}. These effects, along with others like charge-odd directed flow \cite{Gursoy:2014aka, Das:2016cwd, Gursoy:2018yai, Zhang:2022lje, STAR:2023jdd}, are subjects of ongoing researches (see \cite{Huang:2015oca, Hattori:2022hyo, Gao:2020vbh} for reviews). Similarly, The QGP with potent magnetic fields may also exist in the early universe and the core of magnetars \cite{Bali:2011qj, Vachaspati:1991nm, Duncan:1992hi}. For the former, the magnetic field strength might have reached an astonishing $4\text{GeV}^2$, reflecting the severe conditions of the early cosmos. On the other hand, in magnetar cores, the magnetic field strength is generally about  $10^{-5} \text{GeV}^2$.

Heavy quarkonium, the flavor singlet bound state consists of heavy quark and the corresponding anti-quark ($c\bar{c}$, $b\bar{b}$), serves as a probe to demonstrate the existence of QGP in heavy-ion collisions. In recent years, using gauge/gravity duality, extensive works on the properties of heavy quarkonium have been done, including quarkonium spectra \cite{Zhao:2021ogc, Zhao:2023pne, Braga:2023fac, Mamani:2022qnf, Braga:2018zlu, Braga:2017bml, Zhu:2024ddp, Cao:2021tcr}, thermal width \cite{Zhao:2019tjq, Feng:2019boe, Braga:2016oem, Finazzo:2013rqy}, configuration entropy \cite{Zhao:2023yry, Braga:2023qee, Braga:2021fey, Braga:2021zyi, Braga:2020opg}, potential \cite{Chen:2023yug, Zhou:2021sdy, Zhou:2020ssi, Zhang:2023psy, Tahery:2022pzn, Zhang:2011zj, Chu:2009qt, Hou:2007uk}, and other related thermodynamical properties \cite{Hou:2021own, Chen:2017lsf, Chen:2022obe, Zhu:2019igg, Chen:2021gop, Wu:2022ufk, Li:2021otz, Zhu:2023aaq, Sun:2024anu, Wei:2023pdf}. These studies appear to demonstrate a thorough understanding of the properties of heavy quarkonium. However, recent measurements by the STAR collaboration in the RHIC beam energy scan \cite{STAR:2022fan} and by the ALICE collaboration at LHC energies \cite{ALICE:2022dyy} suggest that $\phi$ and $J/\psi$ mesons show a preference for longitudinal and transverse polarization, respectively. This contrasting physical phenomenon places another barrier in scientists' path toward a comprehensive understanding of the properties of vector mesons. Currently, there lacks a robust theoretical framework to explain the distinctions of $\phi$ and $J/\psi$ mesons coherently.

Experimentally, the generation of heavy quarkonia is a two-step process \cite{Braun-Munzinger:2000csl, ALICE:2022nfv}. Firstly, there's the evanescent yet impactful hard process, where perturbative QCD unveils the creation of fleetingly existing heavy quark pairs. Secondly, a more enduring soft process comes into play, artfully described by non-perturbative QCD, capturing the charming interplay of quark pairs as they elegantly bind together over extended periods. The resulting heavy quarkonia, in particular at low transverse momentum $p_T$, might exhibit nontrivial spin alignment concerning a chosen axis \cite{Faccioli:2010kd}, which is also referred to as the polarization for heavy quarkonia. This phenomenon is usually parameterized by the the elements of the spin density matrix \cite{Schilling:1969um, Sheng:2022ssp, Sheng:2022wsy, Sheng:2020ghv, Sheng:2019kmk} or by five $\lambda$-parameters $\{\lambda_{\theta},\, \lambda_\varphi,\,\lambda_{\theta\varphi},\,\lambda_\varphi^\perp,\,\lambda_{\theta\varphi}^\perp\}$; the latter ones are related to the angular distribution of the decay product of quarkonia \cite{Faccioli:2010kd}
\begin{eqnarray} \label{angular-distribution}
    W(\theta^\ast,\varphi^\ast)&\propto& \frac{1}{3+\lambda_\theta}\left(1+\lambda_{\theta}\text{cos}^2\theta^\ast+\lambda_\varphi \sin^2\theta^\ast\cos2\varphi^\ast+\lambda_{\theta\varphi}\sin2\theta^\ast\cos\varphi^\ast\right. \nonumber\\
   && \left.+\lambda_\varphi^\perp \sin^2\theta^\ast\sin2\varphi^\ast+\lambda_{\theta\varphi}^\perp \sin2\theta^\ast\sin\varphi^\ast\right)
\end{eqnarray}
where $\theta^\ast$ and $\varphi^\ast$ are the polar angle and azimuthal angle of one decay daughter in the rest frame of the mother particle. For dilepton decays with the lepton's mass much smaller than the meson, these parameters are related to the elements of spin density matrix as
\begin{eqnarray}\label{lambda-parameters}
   && \lambda_\theta=\frac{1-3\rho_{00}}{1+\rho_{00}},\  \lambda_\varphi=\frac{2\text{Re}\rho_{1,-1}}{1+\rho_{00}},\ \lambda_{\theta\varphi}=\frac{\sqrt{2}\text{Re}(\rho_{01}-\rho_{0,-1})}{1+\rho_{00}},\nonumber\\
   &&\lambda_\varphi^\perp=\frac{-2\text{Im}\rho_{1,-1}}{1+\rho_{00}},\ \lambda_{\theta\varphi}^\perp=\frac{\sqrt{2}\text{Im}(\rho_{01}+\rho_{0,-1})}{1+\rho_{00}}.
\end{eqnarray}
where $\lambda_\theta>0$ and $\lambda_\theta<0$ correspond to transversely and longitudinally polarized states, respectively.

Recently, the ALICE collaboration measured the polarization of $J/\psi$ and $\Upsilon(1S)$ through the dimuon channel in Pb-Pb collisions. The measurements were carried out for $J/\psi$ and $\Upsilon(1S)$ in both the helicity frame and Collins-Soper frame, respectively  \cite{ALICE:2020iev}, and also for $J/\psi$ in the direction of event plane. \cite{ALICE:2022dyy}. The results suggest a transverse polarization in the helicity frame for the $J/\psi$. Considering the fact that the initial hard process plays an important role in the $J/\psi$ production, we expect that the strong magnetic field during the initial stage of heavy-ion collisions has a sizable effect on the polarization of $J/\psi$. Due to the strong interaction in the non-perturbative regime, a powerful non-perturbative tool becomes particularly important.

In this study, we employ gauge/gravity duality to explore the spin alignment for $J/\psi$ meson within a magnetized quark-gluon plasma (QGP). Our study is particularly focused on the effects of intense magnetic fields on the spin alignment of $J/\psi$ meson. We model realistic QGP conditions using a five-dimensional soft wall framework, wherein a significant magnetic field is introduced through the U(1) gauge field. Our approach involves calculating the spectral function and determining the effective mass of $J/\psi$ mesons for various spin states. We then scrutinize the influence of the colossal magnetic field on spin alignment across different spin quantization directions.

The article is organized as follows: Section \ref{sec:01} lays out our holographic model and defines the two-point correlation function. In Section \ref{sec:02}, we delve into numerical results for the spectral function and polarization of the $J/\psi$ meson, considering two distinct cases: one with the magnetic field parallel to the $J/\psi$ momentum and the other with the magnetic field perpendicular to it. Section \ref{sec:03} applies our model to real heavy-ion collisions. Finally, Section \ref{sec:04} offers a concise summary of our paper.

Since the density matrix and the $\lambda$ parameters depend on the choice of spin quantization direction. In this paper, we label the quantities in the helicity frame with a script "H", those in the Collins-Soper frame with "CS", and the quantities in the case with spin quantization direction along the direction of event plane with "EP".

\section{Holographic model}\label{sec:01}
\subsection{Set up of background}\label{sec:01_model}
We use a 5-dimensional soft wall model to describe the QGP environment.  The action takes the following form \footnote{Here $\gamma$ is the determinant of the induced metric $\gamma_{\mu\nu}$ at $\zeta\rightarrow \infty$ and $K:=\gamma^{\mu\nu}K_{\mu\nu}=-\sqrt{g^{\zeta\zeta}}\frac{\partial_\zeta\sqrt{\gamma}}{\sqrt{\gamma}}$ denotes the trace of the extrinsic curvature where the extrinsic curvature is defined by the outward point normal vector to the boundary.}~\cite{Dudal:2015wfn}:
\begin{align}
    S &=\frac{1}{16\pi G_5}\int d^5x\sqrt{-g}\left(R+\frac{12}{L^2}\right)+\frac{1}{8\pi G_5}\int d^4x\sqrt{-\gamma}\left(K-\frac{3}{L}\right)+S_{\text{f}}\label{eq1},\\
    S_{\text{f}} &=-\frac{N_c}{16\pi^2}\int d^4x\int_0^{\zeta_h} d\zeta Q(\zeta)\text{Tr}\left(F_L^2+F_R^2 \right)\label{eq2},
\end{align}
Here $Q(\zeta)=\frac{e^{-\phi}\sqrt{-g}}{4g_5^2} $, with $g_5$ representing Yang-Mills coupling constant. This coupling constant is determined by matching the UV asymptotics of the current-current two-point function between bulk and boundary theories, $g_5^2=\frac{12\pi^2}{N_c}$, as detailed in \cite{Erlich:2005qh}. Additionally, in holographic theory, one primarily focuses on the 't Hooft coupling constant given as $\lambda=g_{5}^{2}N_{c}$ (where $N_{c}$ refers to the color number) which is fixed in large $N_c$ limit. In Eq.\eqref{eq1}, the first term is the Einstein-Hilbert action, the second term is the boundary action consisting of the Gibbons-Hawking surface term and the last term is the flavor action including a soft wall dilaton which ensures the IR effective cut-off of the theory. The dilaton takes the standard choice~\cite{Karch:2006pv}, $\phi=c_i \zeta^2$, where the parameter $c_i$ is directly related to the meson's mass spectrum and $i$ represents different mesons. For background, we only consider the up and down flavors, i.e. the parameter $c_\rho=0.151\, \text{GeV}^2$ is fixed by the $\rho$ meson mass (linear Regge behavior) with the scaling dimension $\Delta=3$ (corresponding to the bulk mass $m_5^2L^2=-3$). It is important to note that authors of \cite{Dudal:2015wfn} have assumed that the dilaton field does not significantly backreact on the metric to analyze the Hawking-Page transition using the soft wall model, i.e. it does not affect the gravitational dynamics(see \cite{Dudal:2015wfn} for more details). By setting $F_L=F_R\sim \mathfrak{B}$(5-dimensional magnetic field), we can get the following perturbation solution~\cite{DHoker:2009mmn,DHoker:2009ixq,Dudal:2015wfn}:
\begin{equation}\label{eq3}
    ds^2=\frac{L^2}{\zeta^2}\left(-f(\zeta)dt^2+h_T(\zeta)(dx^2+dy^2)+h_P(\zeta)dz^2+\frac{d\zeta^2}{f(\zeta)}\right).
\end{equation}
with \footnote{One can find that the blackening coefficients between \cite{Dudal:2015wfn} and \cite{Braga:2021fey} could be connected through setting $x=\frac{1}{yz_h}$. To ensure that the perturbative solution is meaningful, in principle, it is sufficient to ensure that the perturbation terms in Eq.\eqref{eq4} are smaller than at least one of the first two terms, as detailed in \cite{Dudal:2015wfn}. The logarithm itself is usually $\mathcal{O}(1)$. Therefore, we can obtain $eB<\sqrt{\frac{3}{2}}\frac{1.6}{\zeta_h^2}\approx\frac{1.96}{\zeta_h^2}$.}
\begin{align}
  f(\zeta) &= 1-\frac{\zeta^4}{\zeta_h^4}+\frac{2}{3}\frac{e^2B^2}{1.6^2}\zeta^4\ln{\frac{\zeta}{\zeta_h}}+\mathcal{O}(e^4B^4)\label{eq4},\\
  h_T(\zeta) &= 1-\frac{4}{3}\frac{e^2B^2}{1.6^2}\zeta_h^4\int_{0}^{\zeta/\zeta_h}\frac{y^3\ln{y}}{1-y^4}dy+\mathcal{O}(e^4B^4)\label{eq5},\\
  h_P(\zeta) &= 1+\frac{8}{3}\frac{e^2B^2}{1.6^2}\zeta_h^4\int_{0}^{\zeta/\zeta_h}\frac{y^3\ln{y}}{1-y^4}dy+\mathcal{O}(e^4B^4)\label{eq6},
\end{align}
where $\zeta$ is the holographic radial coordinate with the asymptotic AdS boundary at $\zeta\rightarrow0$. The event horizon $\zeta_h$ is determined by $f(\zeta_h)=0$.  Here we introduce the constant magnetic field in the $z$-direction, which breaks the SO(3) rotation symmetry. Note that $e{B}$ denotes a 4-dimensional physical magnetic field related to the 5-dimensional magnetic field $\mathfrak{B}$ by $e{B}=\frac{1.6}{L}\mathfrak{B}$ (see more details in \cite{Dudal:2015wfn}). The Hawking temperature can be written as
\begin{equation}\label{eq7}
    T=\frac{1}{4\pi}\bigg|\frac{4}{\zeta_h}-\frac{2}{3}\frac{e^2B^2}{1.6^2}\zeta_h^3\bigg|.
\end{equation}

\subsection{Holographic description of \texorpdfstring{$J/\psi$}{Jpsi} meson}\label{sec:01_meson}
Given the well-constructed background, we will consider placing vector mesons in the background to study the impact of the background on it. It is noted that the vector mesons do not back reaction to the background. We take the generic Maxwell action to describe the vector mesons,
\begin{equation}\label{eq8}
  S_M=-\int d^4xd\zeta Q(\zeta) \left[F_{MN}F^{MN} \right],
\end{equation}
where $F_{MN}=\nabla_M A_N-\nabla_N A_M=\partial_M A_N-\partial_N A_M$ with $\nabla_M$ representing the covariant derivative and $Q(\zeta)$ has been given in subsection \ref{sec:01_model}. Note that our study focuses not on the direct impact of the magnetic field on the vector meson, but rather on its effects on the background plasma. Consequently, while the metric in our model is dependent on $eB$, the Lagrangian of vector meson remains unaffected (The justification for magnetic field not directly entering in this meson action is shown in the appendix \ref{App1}).  Here we consider $J/\psi$ meson, therefore the mass parameter is expressed by $c_{J/\psi}$. According to the KKSS model~\cite{Karch:2006pv}, linear $m^2$ spectrum obeys such a constraint $m_{n,S}^2=4c_{J/\psi}(n+S)$ with spin $S$ and radial excitation number $n$ ($n=0,1,2,3,\cdots$). As we are only interested in $J/\psi$ meson, we fix the parameter $c_{J/\psi}$ solely based on the mass of $J/\psi$ meson. One can easily obtain $c_{J/\psi}=m_{J/\psi}^2/4$ with $m_{J/\psi}\approx3.097\, \text{GeV}$ from the Particle Data Group (PDG) ~\cite{ParticleDataGroup:2018ovx}.

By performing a variational analysis of the generic Maxwell action Eq.\eqref{eq8}, we can derive the equation of motion, which is as follows:
\begin{equation}\label{eq9}
  \partial_M\left(Q(\zeta)F^{MN}\right)=0.
\end{equation}
Here we take the radial gauge $A_\zeta=0$ and consider the Fourier transform
\begin{equation}\label{eq10}
  A_\mu(x,\zeta)=\int \frac{d^4 p}{(2\pi)^4}e^{-i\omega t+i{\bf p}\cdot {\bf x}}A_\mu(p,\zeta),
\end{equation}
where $p$ labels the four-momentum $p_\mu=(p_t,{\bf p})=(-\omega, p_x,p_y,p_z) $. With the help of EOM~\eqref{eq9} and Eq.~\eqref{eq10}, the action ~\eqref{eq8} can be expressed as
\begin{equation}\label{eq11}
 S_M=-2\int \frac{d^4p}{(2\pi)^4}Q(\zeta)g^{\zeta\zeta}g^{\mu\nu}A_\mu(-p,\zeta)\partial_\zeta A_\nu(p,\zeta)\bigg|_0^{\zeta_h}.
\end{equation}
Furthermore, by introducing gauge invariant electric field,
\begin{equation}
E_{i}(p,\zeta)\equiv-p_t A_{i}(p,\zeta)+p_{i}A_{t}(p,\zeta)
\end{equation}
the action (\ref{eq11}) can be rewritten as
\begin{equation}\label{eq12}
 S_M=-2\int \frac{d^4p}{(2\pi)^4}Q(\zeta)\frac{g^{\zeta\zeta}}{p_0} g^{ik}G_k^j E_i(-p,\zeta)\partial_\zeta E_j(p,\zeta)\bigg|_0^{\zeta_h}
\end{equation}
with $G_k^j=(-g_k^j+p_kp_ig^{ij}/p^2)/p_t$, where $p^2$ represents the square of four-momentum $p^2=p_\mu p^\mu$, $g_k^j=g^{ij}\delta_{ki}$ and $\delta_{ki}$ is the Kronecker delta. To obtain the correlation functions, it is beneficial to decompose the electric field as the product of two functions. One of these functions, denoted as $\widetilde{E}_{i}(e_{j},\zeta,p)$, solely depends on the holographic coordinate and satisfies the boundary condition $\lim_{\zeta\rightarrow0}\widetilde{E}_{i}(e_{j},\zeta,p)=\delta_{ij}$, while the other function depends on the boundary value $E^0_k(q)=\lim_{\zeta\rightarrow0}E_i(q,\zeta)$
\begin{equation}\label{eq13}
E_i(p,\zeta)=\widetilde{E}_{i}(e_{j},\zeta,p)E^0_j(p),\quad (j=x, y, z)
\end{equation}
Then the action can be expressed in the form
\begin{equation}\label{eq14}
 S_M=\int \frac{d^4p}{(2\pi)^4}E_l^0(-p)H^{ls}E_s^0(p), \quad H^{ls}=-2Q(\zeta)\frac{g^{\zeta\zeta}}{p_t}g^{ik}G_k^j\widetilde{E}_{i}(e_{l},\zeta,-p)\partial_\zeta\widetilde{E}_{j}(e_{s},\zeta,p).
\end{equation}
Finally, we return to the gauge field using $E_l^0(\pm q)=\pm C_l^\mu A_\mu^0(\pm q)$. The action is rewritten as
\begin{equation}\label{eq15}
 S_M=\int \frac{d^4p}{(2\pi)^4}A_\mu^0(-p)\mathcal{F}^{\mu\nu}A_\nu^0(p), \quad \mathcal{F}^{\mu\nu}=-C_l^\mu H^{ls} C_s^\nu.
\end{equation}
The current-current correlation function using the Son-Starinets prescription \cite{Son:2002sd} can be defined as \footnote{Note that this definition only applies to cases where space-time geometry is asymptotic AdS.}
\begin{equation}\label{eq16}
 D^{\mu\nu}(p)=2\lim_{\zeta\to0}\mathcal{F}^{\mu\nu}(\zeta,p).
\end{equation}
The spectral function can be obtained by the imaginary part of the correlation function. It is easy to verify that  satisfies that $D^{\mu\nu}$ obeys the Ward identity $p_\mu D^{\mu\nu}(p)=p_\mu D^{\nu\mu}(p)=0$, which allows us to decompose it  as
\begin{equation} \label{decompose-propagator}
\text{Im} D^{\mu\nu}(p)=-\sum_{\lambda\lambda^\prime}v^\mu(\lambda,p)v^{\ast\nu}(\lambda^\prime,p)\varrho_{\lambda\lambda^\prime}(p)
\end{equation}
where the polarization vectors are defined as
\begin{equation}\label{eq18}
 v^{\mu}(\lambda,p)=\left(-\frac{{\bf p}\cdot\boldsymbol\epsilon_\lambda}{M}, \boldsymbol\epsilon_\lambda+\frac{{\bf p}\cdot\boldsymbol\epsilon_\lambda}{M(\omega+M)}{\bf p} \right)
\end{equation}
with $\boldsymbol\epsilon_\lambda$ being the spin 3-vector in the meson's rest frame. Here $M\equiv\sqrt{\omega^2-{\bf p}^2}$ is the invariant mass. One can check that $v^\mu(\lambda,p)$ satisfy the orthonormality conditions $\eta_{\mu\nu}v^{\ast\mu}(\lambda,p)v^\nu(\lambda^\prime,p)=\delta_{\lambda\lambda^\prime}$  and the completeness condition $\sum_{\lambda} v^\mu(\lambda,p)v^{\ast\nu}(\lambda,p)=(\eta^{\mu\nu}-p^\mu p^\nu/p^2)$. We then identify $\varrho_{\lambda\lambda^\prime}$ as the spectral function in the spin space, which can be extracted from $D^{\mu\nu}$ as
\begin{equation} \label{spectral-spin}
\varrho_{\lambda\lambda^\prime}(p)=-\eta_{\mu\alpha}\eta_{\nu\beta}v^{\ast\mu}(\lambda,p)v^\nu(\lambda^\prime,p)\text{Im} D^{\alpha\beta}(p)
\end{equation}
We note that the spectral function in Eq. (\ref{spectral-spin}) depends on the choice of spin quantization direction $\boldsymbol\epsilon_0$. For example, the choice $\boldsymbol\epsilon_0={\bf p}/|{\bf p}|$  is referenced as the helicity frame in literature. Other choices for $\boldsymbol{\epsilon}_0$  include the event-plane direction $\boldsymbol{\epsilon}_0=(0,1,0)$, and the Collins-Soper frame. Vectors $\boldsymbol{\epsilon}_1$ and $\boldsymbol{\epsilon}_{-1}$ are determined to be orthogonal to $\boldsymbol{\epsilon}_0$ and therefore they also depend on the choice of $\boldsymbol{\epsilon}_0$.

\subsection{Equations of motion}\label{sec:01_eom}
Through Eq.~\eqref{eq14}, a crucial insight arises: to achieve numerical solutions for correlation function, a prerequisite is an acquaintance with the holographic function $\partial_\zeta\widetilde{E}_{j}(e_{s},\zeta,p)$ derived by effectively solving the equation of motion ~\eqref{eq9}. We now study the equations of motion~\eqref{eq9}. The component equations of motion are as follows:
\begin{align}\label{eq22}
A_t^{\prime\prime}+\left(-\frac{1}{\zeta}+\frac{h_T^\prime}{h_T}+\frac{h_P^\prime}{2h_P}-\phi^\prime\right)A_t^\prime-\left(\frac{p_z\left(p_zA_t+\omega  A_z\right)}{fh_P}+\frac{\left(p_x^2+p_y^2\right)A_t+\omega\left(p_xA_x+ p_y A_y\right)}{fh_T}\right)&=0,\nonumber\\
A_x^{\prime\prime}+\left(-\frac{1}{\zeta}+\frac{f^\prime}{f}+\frac{h_P^\prime}{2h_P}-\phi^\prime\right)A_x^\prime+\left(\frac{\omega\left(p_xA_t+\omega  A_x\right)}{f^2}+\frac{p_y\left(p_xA_y-p_yA_x\right)}{fh_T}+\frac{p_z\left( p_xA_z- p_z A_x\right)}{fh_P}\right)&=0,\nonumber\\
A_y^{\prime\prime}+\left(-\frac{1}{\zeta}+\frac{f^\prime}{f}+\frac{h_P^\prime}{2h_P}-\phi^\prime\right)A_y^\prime+\left(\frac{\omega\left(p_yA_t+\omega  A_y\right)}{f^2}+\frac{p_x\left(p_yA_x-p_xA_y\right)}{fh_T}+\frac{p_z\left( p_yA_z- p_z A_y\right)}{fh_P}\right)&=0,\nonumber\\
A_z^{\prime\prime}+\left(-\frac{1}{\zeta}+\frac{f^\prime}{f}+\frac{h_T^\prime}{h_T}-\frac{h_P^\prime}{2h_P}-\phi^\prime\right)A_z^\prime+\left(\frac{p_z\left(p_xA_x+p_y A_y\right)-(p_x^2+p_y^2)A_z}{fh_T}+\frac{\omega\left(p_zA_t+\omega A_z\right)}{f^2}\right)&=0,\nonumber\\
 fh_P\left(p_xA_x^\prime+p_yA_y^\prime\right)+h_T\left(\omega h_PA_t^\prime+p_zfA_z^\prime \right)&=0,
 \end{align}
where the prime ($^\prime$) represents the derivative with respect to $\zeta$. These five equations describe the coupling between the gauge field fluctuation and the background metric involving the magnetic field. Note that the functions $h_T,h_P$ refer to the metric \eqref{eq3} including the magnetic field. The corresponding electric field component, $E_j=\omega A_j+ q_j A_t$, equations are as follows,
\begin{align}\label{eq23}
E_x^{\prime\prime}+\left(-\frac{1}{\zeta}+\frac{f^\prime}{f}+\frac{h_P^\prime}{2h_P}-\phi^\prime+p_x^2\Xi_T\right)E_x^\prime+p_xp_y\Xi_T E_y^\prime+p_xp_zh_T\Xi_T E^\prime_z+\Theta E_x&=0,\nonumber\\
E_y^{\prime\prime}+\left(-\frac{1}{\zeta}+\frac{f^\prime}{f}+\frac{h_P^\prime}{2h_P}-\phi^\prime+p_y^2\Xi_T\right)E_y^\prime+p_xp_y\Xi_T E_x^\prime+p_yp_zh_T\Xi_T E^\prime_z+\Theta E_y&=0,\nonumber\\
E_z^{\prime\prime}+\left(-\frac{1}{\zeta}+\frac{f^\prime}{f}+\frac{h_T^\prime}{h_T}-\frac{h_P^\prime}{2h_P}-\phi^\prime+p_y^2\Xi_P\right)E_z^\prime+p_z\frac{h_P}{h_T}\Xi_P\left(p_x E_x^\prime+p_y E^\prime_y\right)+\Theta E_z&=0,
 \end{align}
with
\begin{align}\label{eq24}
    \Theta &=\frac{\omega^2}{f^2}-\frac{p_x^2+p_y^2}{fh_T}-\frac{p_z^2}{fh_P},\quad
     \Xi_T=\frac{h_P \left(fh_T^\prime-h_Tf^\prime\right)}{h_T\left(f\left(p_z^2h_T+\left(p_x^2+p_y^2\right)h_P\right)-\omega^2h_Ph_T\right)},\nonumber\\
    \Xi_P &=\frac{h_T \left(fh_P^\prime-h_Pf^\prime\right)}{h_P\left(f\left(p_z^2h_T+\left(p_x^2+p_y^2\right)h_P\right)-\omega^2h_Ph_T\right)}.
\end{align}
Next, we will provide a general method for solving this type of second-order ordinary differential equation system\footnote{The similar method is taken in \cite{Mamani:2013ssa, Mamani:2022qnf}. They consider the wave vector $p_\mu=(-\omega,0,0,p_z)$, and the corresponding equations of motion are relatively simple.}. The system of equations ~\eqref{eq23} can be rewritten as follows:
\begin{equation}\label{eq25}
    [\delta_{ij}\partial_\zeta^2 +K_{ij}\partial_\zeta+\Theta \delta_{ij}]E_j(\zeta)=0,
\end{equation}
where $K_{ij}$, $i,j=x,y,z,$ is a matrix function of $\zeta$.

Near the horizon, we impose the incoming wave solution as the initial condition, therefore we have
\begin{equation}\label{eq26}
    E(\zeta)=e^{-i\omega r_\ast}\psi(\zeta),
\end{equation}
where $r_\ast$ represents the tortoise coordinate defined as $\partial_{r_\ast}=-f(\zeta)\partial_\zeta$. Subsequently, the system of equations ~\eqref{eq25} can be represented using the wave function $\psi(\zeta)$,
\begin{equation}\label{eq27}
    \left[\partial_\zeta^2+\widehat{K}\partial_\zeta+\widehat{N} \right]\psi(\zeta)=0
\end{equation}
with
\begin{equation}\label{eq28}
    \widehat{K}=K+\frac{2i\omega}{f(\zeta)}\mathbb{I}_{3\times3},\quad \widehat{N}=\frac{i\omega}{f(\zeta)}\widehat{K}+\frac{\omega^2-i\omega\partial_\zeta f(\zeta)}{f(\zeta)^2}\mathbb{I}_{3\times3}+\Theta\mathbb{I}_{3\times3}.
\end{equation}
where $\mathbb{I}_{3\times3}$ denotes the $3\times3$ unit matrix. An important point to note is that $\widehat{K}$ and $\widehat{N}$ diverge at the horizon. Therefore, multiplying both sides of the equation by $\left(\zeta/\zeta_h-1\right)^\nu$ proves to be highly useful, as it eliminates the singularity of the equation system. We denote $\widetilde{I}=\left(\zeta/\zeta_h-1\right)^\nu\mathbb{I}_{3\times3}$, $\widetilde{K}=\left(\zeta/\zeta_h-1\right)^\nu\widehat{K}$ and $\widetilde{N}=\left(\zeta/\zeta_h-1\right)^\nu\widehat{N}$ with $\nu$ representing the minimum integer value that guarantees $\widetilde{I}$, $\widetilde{K}$ and $\widetilde{N}$ to remain finite. Then the system of equations ~\eqref{eq23} can be rewritten as
\begin{equation}\label{eq29}
    \left[\widetilde{I}\partial_\zeta^2+\widetilde{K}\partial_\zeta+\widetilde{N} \right] \psi(\zeta)=0.
\end{equation}
It is of utmost importance to emphasize that $\psi(\zeta)$ serves as the present unknown function, being a $3\times1$ matrix. We assume that $\psi(\zeta)$ can be expanded in a Taylor series near the horizon as
\begin{equation}\label{eq30}
    \psi(\zeta)=\varphi+\sum_{n=1}^{\infty}a_n\left(\frac{\zeta}{\zeta_h}-1\right)^n,
\end{equation}
where $\varphi$ represents the value of $\psi(\zeta)$ at the horizon and $a_n$ is a $3\times 1$ matrix. Substituting this solution into equation ~\eqref{eq29}, we obtain
\begin{equation}\label{eq31}
    \left[\widetilde{I}\partial_\zeta^2+\widetilde{K}\partial_\zeta+\widetilde{N} \right] \sum_{n=1}^{\infty}a_n\left(\frac{\zeta}{\zeta_h}-1\right)^n=-\widetilde{N}\varphi.
\end{equation}
For each element in $\widetilde{I}$, $\widetilde{K}$ and $\widetilde{N}$, we also perform a Taylor series expansion, that is
\begin{equation}\label{eq32}
   \widetilde{I}=\sum_{j=0}^{\infty}\widetilde{I}_j\left(\frac{\zeta}{\zeta_h}-1\right)^j,\quad \widetilde{K}=\sum_{l=0}^{\infty}\widetilde{K}_l\left(\frac{\zeta}{\zeta_h}-1\right)^l,\quad
   \widetilde{N}=\sum_{s=0}^{\infty}\widetilde{N}_s\left(\frac{\zeta}{\zeta_h}-1\right)^s.
\end{equation}
Then we have
\begin{align}\label{eq33}
   \sum_{j=0}^\infty\sum_{n=1}^\infty n(n-1)\widetilde{I}_j a_n \left(\frac{\zeta}{\zeta_h}-1\right)^{n+j-2}
   + \sum_{l=0}^\infty\sum_{n=1}^\infty n\widetilde{K}_l a_n \left(\frac{\zeta}{\zeta_h}-1\right)^{n+l-1}\nonumber\\
   +\sum_{s=0}^\infty\sum_{n=1}^\infty\widetilde{N}_s a_n \left(\frac{\zeta}{\zeta_h}-1\right)^{n+s}
    =-\sum_{i=0}^\infty\widetilde{N}_i \varphi \left(\frac{\zeta}{\zeta_h}-1\right)^{i},
\end{align}
which gives the following constraint equation
\begin{equation}\label{eq34}
   \sum_{j=0}^\infty\sum_{n=1}^\infty n(n-1)\widetilde{I}_j a_n \delta_{n+j-2,i}
   + \sum_{l=0}^\infty\sum_{n=1}^\infty n\widetilde{K}_l a_n \delta_{n+l-1,i} +
   \sum_{s=0}^\infty\sum_{n=1}^\infty\widetilde{N}_s a_n \delta_{n+s,i}=-\widetilde{N}_i \varphi,
\end{equation}
where $\delta_{x,i} (x=n+j-2,n+l-1,n+s)$ is the Kroneck delta. Next, we define a matrix $M_{in}$ with its element given by
\begin{equation}\label{eq35}
  M_{in}= \sum_{j=0}^\infty n(n-1)\widetilde{I}_j \delta_{n+j-2,i}
   + \sum_{l=0}^\infty\widetilde{K}_l a_n \delta_{n+l-1,i} +
   \sum_{s=0}^\infty\widetilde{N}_s a_n \delta_{n+s,i},
\end{equation}
where $M_{in}$ is $3\times3$ block matrix and every block matrix has $i\times n$ elements. In addition, we also define $\widetilde{\varphi}_i=-\widetilde{N}_i \varphi$. Then the equation ~\eqref{eq34} can be written as\footnote{Note that the matrix $M_{in}$ in Eq.\eqref{eq36} has an inverse matrix only when $\nu=1$ in Eq.\eqref{eq29}. If $\nu=2$, we find $\widetilde{\varphi}_0=M_{0n}a_n=0$ for all $n=1,2,3,\cdots$. Therefore, the value of $i$ starts from $1$.}
\begin{equation}\label{eq36}
    \widetilde{\varphi}_i=\sum_{n=1}^\infty M_{in}a_n,\quad i=1,2,3,\cdots.
\end{equation}
Finally, we can obtain the unknown quantities
\begin{equation}
    a_n=\sum_{i=1}^\infty(M_{in})^{-1} \widetilde{\varphi}_i.
\end{equation}
Then we can numerically solve Eq.~\eqref{eq23} by taking $a_n$ into Eq.~\eqref{eq30} and ~\eqref{eq26} to obtain the initial condition of the equation, that is, the values of the electric field and its derivative at the event horizon.

\subsection{Dimuon production rate and spin alignment}

In experiments, the spin alignment for $J/\psi$  is measured through the angular distribution of daughter particles in the decay process $J/\psi\rightarrow\mu^+\mu^-$.  Considering the process of an initial state $i$ to a final state $f$ with a dimuon pair $\mu^+\mu^-$, the S-matrix element reads \cite{Gale:1990pn},
\begin{equation} \label{S-matrix}
S_{fi}=\int d^4x\int d^4y \left\langle f,\mu^+\mu^-\left|J_\alpha(y)G_R^{\alpha\beta}(x-y)J_\beta^l(x)\right|i\right\rangle
\end{equation}
where $J_\alpha$ is the current that couples to $J/\psi$, which could be the charm quark current if we consider the quark-gluon plasma as the initial state. On the other hand, $J_\beta^l$  is the leptonic current,
\begin{equation}
J_\beta^l(x)\equiv g_{M\mu^+\mu^-}\overline{\psi}_l(x)\Gamma_\beta\psi_l(x)
\end{equation}
where $\Gamma_\beta$ is the effective vertex for the coupling between $J/\psi$ and $\mu^+\mu^-$, with  $g_{M\mu^+\mu^-}$ being the coupling strength, and $\psi_l$ is the lepton field (muon field in this paper). In Eq. (\ref{S-matrix}), we treat the dimuon production through $J/\psi$ decay as a two-step process. First, the $J/\psi$ meson is produced from the initial state at the spacetime point $y$. Then the meson propagates from the point $y$ to the point $x$ and decay at $x$. The propagator $G_R^{\alpha\beta}$ in Eq. (\ref{S-matrix}) is assumed to the retarded propagator in vacuum, which is given by
\begin{equation} \label{propagator-free}
G^{\alpha\beta}_R(p)=-\frac{\eta^{\alpha\beta}+p^\alpha p^\beta/p^2}{p^2+m_{J/\psi}^2+im_{J/\psi}\Gamma}
\end{equation}
where $m_{J/\psi}=3.097$ GeV is the resonance mass and $\Gamma=0.1$ GeV is the phenomenalogical width for a $J/\psi$ passing though the QGP.
The transition probability for the dimuon production is then given by $R_{fi}\equiv|S_{fi}|^2/(TV)$. After summing over all possible final state $f$ and averaging over the initial state $i$, we obtain the total dimuon production rate
\begin{eqnarray}
N&=&-2g^2_{M\mu^+\mu^-}\int\frac{d^3{\bf p}_+}{(2\pi)^3E_+}\frac{d^3{\bf p}_-}{(2\pi)^3E_-}
[G_R^{\rho\sigma}(p)]^\ast G_R^{\alpha\beta}(p) \nonumber\\
&&\times n_B(\omega)\text{Im} D_{\rho\alpha}(p)\left[p_\beta^-p_\sigma^++p_\beta^+p_\sigma^--g_{\beta\sigma}(p_+\cdot p_-+m_\mu^2)\right]
\end{eqnarray}
where $p^\pm$ are four-momentum for $\mu^\pm$, respectively. The produced muons are assumed to be on-shell, indicating that the energies $E_\pm\equiv\sqrt{{\bf p}^2_\pm+m_\mu^2}$, with $m_\mu=0.105$ GeV is the mass of muon.  We can further express ${\bf p}_\pm$ in terms of the four-momentum of the dimuon pair $p=p^++p^-$ and the polar angle $\theta^\ast$ and azimuthal angle $\varphi^\ast$ for the produced $\mu^+$ in the rest frame of the dimuon pair. The differential production rate is given by
\begin{eqnarray}
\frac{dN}{d^4pd\cos\theta^\ast d\varphi^\ast}&=&-\frac{g^2_{M\mu^+\mu^-}}{2(2\pi)^6}\sqrt{1+\frac{4m_\mu^2}{p^2}}
[G_R^{\rho\sigma}(p)]^\ast G_R^{\alpha\beta}(p) \nonumber\\
&&\times n_B(\omega)\text{Im} D_{\rho\alpha}(p)\left(p_\beta p_\sigma-4 q_\beta q_\sigma-g_{\beta\sigma}p^2\right)
\end{eqnarray}
where $q$  denotes the average of relative momentum between the produced $\mu^+$ and $\mu^-$, and $n_B(\omega)=1/(e^{\omega/T}-1)$  is the Bose-Einstein distribution at temperature $T$. We further substitute Eqs. (\ref{decompose-propagator}) and (\ref{propagator-free}) into the production rate and we are then able to express the result in terms of the spin states,
\begin{eqnarray} \label{dilepton-production}
\frac{dN}{d^4pd\cos\theta^\ast d\varphi^\ast}&=&\frac{3C_N}{8\pi}\left\{\frac{p^2}{p^2-2m_\mu^2}-\frac{p^2+4m_\mu^2}{p^2-2m_\mu^2}\left[\frac{1-\rho_{00}}{2}+\frac{3\rho_{00}-1}{2}\cos^2\theta^\ast\right.\right. \nonumber\\
&&-\text{Re}\rho_{1,-1}\sin^2\theta^\ast\cos2\varphi^\ast+\frac{\text{Re}(\rho_{0,-1}-\rho_{01})}{\sqrt{2}}\sin2\theta^\ast \cos\varphi^\ast\nonumber\\
&&\left.\left.+\text{Im}\rho_{1,-1}\sin^2\theta^\ast\sin2\varphi^\ast-\frac{\text{Im}(\rho_{0,-1}+\rho_{01})}{\sqrt{2}}\sin2\theta^\ast\sin\varphi^\ast\right]\right\}
\end{eqnarray}
where the spin matrix for the produced dimuon is defined as
\begin{equation}
\rho_{\lambda\lambda^\prime}(p)\equiv -\frac{2g^2_{M\mu^+\mu^-}}{3(2\pi)^5C_N}\left(1-\frac{2m_\mu^2}{p^2}\right)\sqrt{1+\frac{4m_\mu^2}{p^2}}\frac{p^2n_B(\omega)\varrho_{\lambda\lambda^\prime}}{(p^2+m_{J/\psi}^2)^2+m_{J/\psi}^2\Gamma^2}
\end{equation}
with $C_N$ being the normalization factor that ensures the matrix is properly normalized as $\sum_{\lambda=0,\pm1}\rho_{\lambda\lambda}=1$. Noting that the invariant mass square for the dilepton pair is restricted near the resonance mass of $J/\psi$, i.e., $p^2\approx -m_{J/\psi}^2$, which is much larger than $m_\mu^2$, we can approximate $m_\mu^2/p^2\approx 0$. By comparing Eq. (\ref{dilepton-production}) with Eq. (\ref{angular-distribution}), we can reproduce the $\lambda$-parameters given in Eq. (\ref{lambda-parameters}), which is a function of the four-momentum $p^\mu=(\omega,{\bf p})$.

\section{Numerical results}\label{sec:02}
In this section, we focus on the effect of magnetic field to the spin alignment for $J/\psi$ mesons with finite momentum.

\subsection{Magnetic field parallel to momentum}

\begin{figure}
  \centering
  \includegraphics[width=0.9\textwidth]{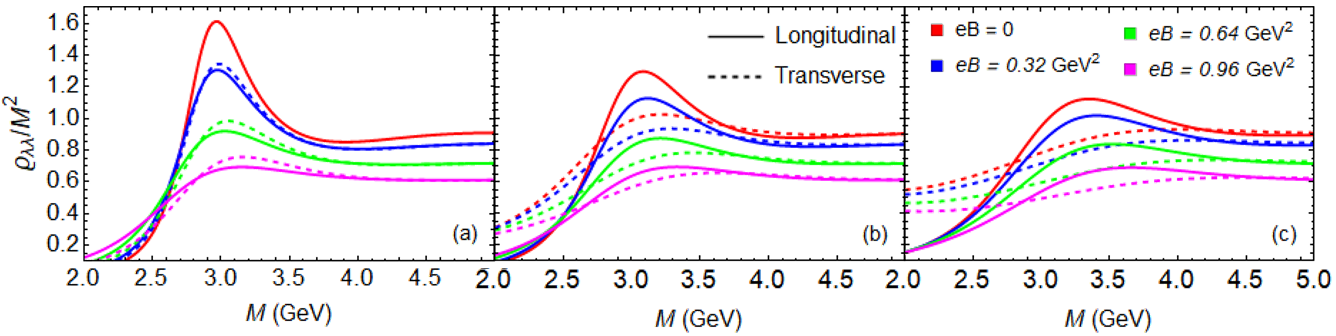}

  \caption{Spectral functions of charmonium, non-dimensionalized by dividing $M^2$, at $T=0.2$ GeV as functions of invariant mass $M$, at momentum $p=0$ (left), $p=5$ GeV (middle), and $p=10$ GeV (right).}\label{fig:spectral-parallel}
\end{figure}

\begin{figure}
  \centering
  \includegraphics[width=0.6\textwidth]{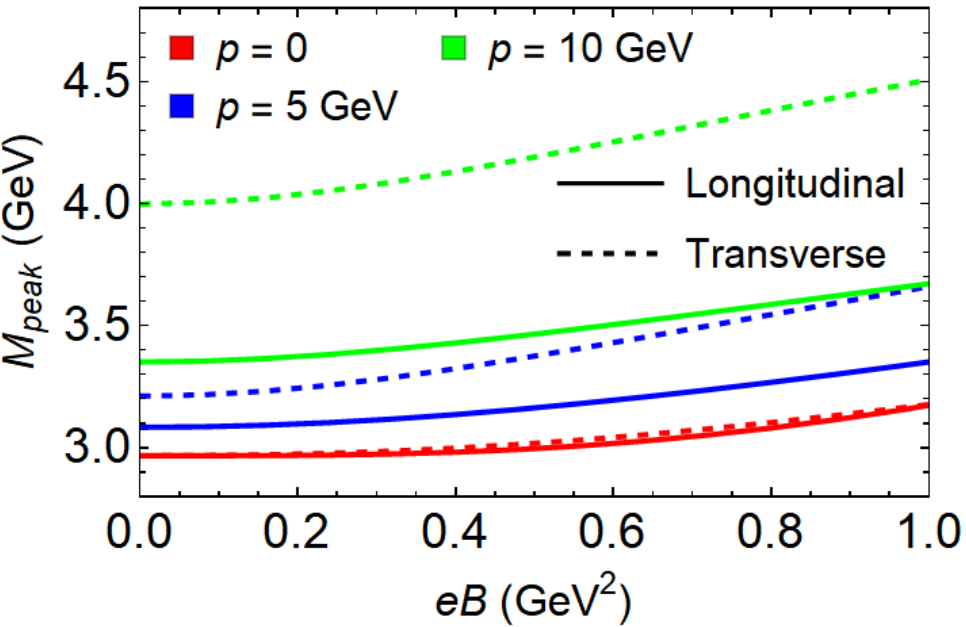}

  \caption{Resonance masses at $T=0.2$ GeV as functions of the magnetic field strength $eB$, for $J/\psi$ mesons with momentum $p=0$ (red lines), $p=5$ GeV (blue lines), and $p=10$ GeV (green lines).}\label{fig:Mpeak-parallel}
\end{figure}

We first consider the case that the magnetic field is parallel to ${\bf p}$. Without loss of generality, we set the magnetic field along the $z$-direction and also set ${\bf p}=(0,0,p)$. In the helicity frame, it's easy to verify that only the diagonal elements of the spectral function in Eq. (\ref{spectral-spin}) have nonvanishing values,
i.e., ${\varrho}_{11}^\text{H}={\varrho}_{-1,-1}^\text{H}\neq 0$ and ${\varrho}_{00}^\text{H}\neq0$, while all off-diagonal elements are zeros.  The symmetry between the longitudinal ($\lambda=0$) mode and transverse ($\lambda=\pm 1$) modes is broken because of the magnetic field or finite momentum, while the two transverse modes are degenerate. We plot in Fig. \ref{fig:spectral-parallel} spectral functions as functions of meson's invariant mass for $J/\psi$ with momentum $p=0$, 5 GeV and 10 GeV at temperature $T=0.2$ GeV. The spectra for longitudinally (transversely) polarized modes are shown by solid (dashed) lines. We observe that a nonzero magnetic field or a nonzero momentum will induce a separation between two different modes. Spectral functions at a relatively large invariant mass, such as $M=3$ GeV, are suppressed by the magnetic field and the nonzero momentum, while spectral functions at a smaller $M$, such as $M=2$ GeV is enhanced. We identify the location of peaks in Fig. \ref{fig:spectral-parallel} as resonance masses for $J/\psi$ mesons. We show in Fig. \ref{fig:Mpeak-parallel} the masses as functions of magnetic field strength. When the meson's momentum is zero, we find that the mass increases with increasing magnetic field strength, while the longitudinally polarized state and transversely polarized states have nearly the same mass, as shown by red solid and red dashed lines in Fig. \ref{fig:Mpeak-parallel}. A nonzero momentum will induce a significant difference between longitudinal and transverse modes, as indicated by blue and green lines.

\begin{figure}
  \centering
  \includegraphics[width=0.6\textwidth]{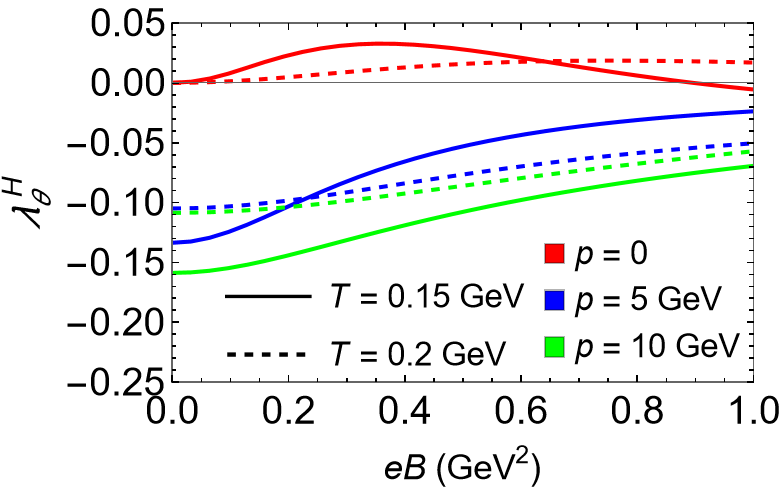}

  \caption{The $\lambda_\theta^\text{H}$ parameter as a function of the magnetic field strength, at momentum $p=0$ (red lines), $p=5$ GeV (blue lines), and $p=10$ GeV (green lines), calculated at temperature $T=0.15$ GeV (solid lines) and $0.2$ GeV (dashed lines).}\label{fig:lambda-theta-B}
\end{figure}

\begin{figure}
  \centering
  \includegraphics[width=0.8\textwidth]{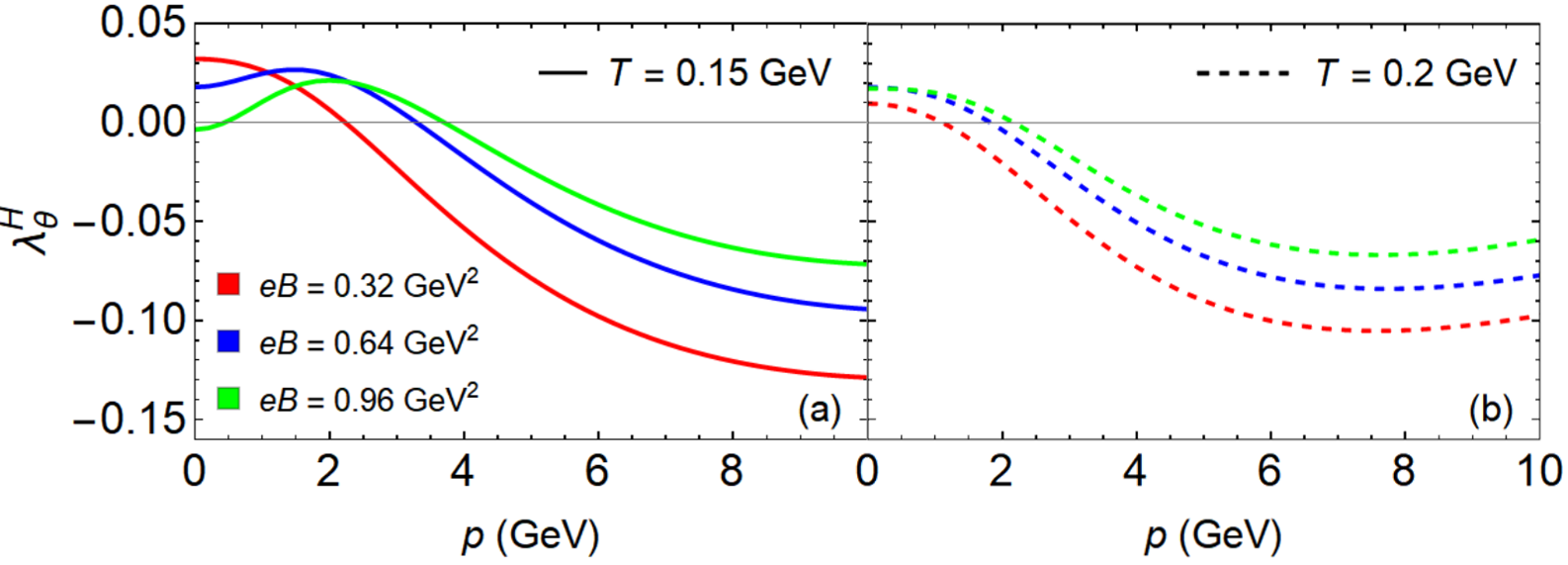}

  \caption{The $\lambda_\theta$ parameter as a function of momentum at $eB=0.32 \text{ GeV}^2$ (red lines), $0.64 \text{ GeV}^2$ (blue lines), and $0.96 \text{ GeV}^2$ (green lines), calculated at temperature $T=0.15$ GeV (left panel) and $0.2$ GeV (right panel).}\label{fig:lambda-theta-p}
\end{figure}

For the case in this section, the only nonvanishing $\lambda$-parameter in the helicity frame is $\lambda_\theta^\text{H}$. In Fig. \ref{fig:lambda-theta-B}, we plot $\lambda_\theta^\text{H}$ as a function of the magnetic field strength. We take two typical temperatures, $T=0.15$ GeV and $T=0.2$ GeV. For static $J/\psi$, the $\lambda_\theta^\text{H}$ parameter is a positive value at both $T=0.15$ GeV and $T=0.2$ GeV, as shown by red lines in Fig. \ref{fig:lambda-theta-B}. At $T=0.2$ GeV, $\lambda_\theta^\text{H}$ increases with increasing field strength for all three cases of momentum. But at $T=0.15$ GeV and $p=0$, $\lambda_\theta^\text{H}$ has a non-monotonic behaviour.  It increases with increasing field strength when $eB<0.32 \text{ GeV}^2$, then decreases when $eB>0.32\text{ GeV}^2$. For the case of $p=0$, we choose the direction of the magnetic field as the spin quantization direction, given that the helicity frame is not well-defined in this scenario. The momentum dependence of $\lambda_\theta^\text{H}$ at difference field strengths are shown in Fig. \ref{fig:lambda-theta-p}. At $T=0.15$ GeV with magnetic field $eB=0.32 \mathrm{GeV}^2$, the $\lambda_\theta^\text{H}$ is positive for a static $J/\psi$. It decreases and reaches a negative value when $p>2.3$ GeV. In a stronger magnetic field, the momentum dependence of $\lambda_\theta^\text{H}$ becomes non-monotonic. As shown by blue and green curves in Fig. \ref{fig:lambda-theta-p} (a), $\lambda_\theta^\text{H}$ first increases and then decreases with $p$. Such a non-monotonic behavior does not exist at a higher temperature $T=0.2$ GeV, as shown in Fig. \ref{fig:lambda-theta-p} (b), where $\lambda_\theta^\text{H}$ decreases as $p$ grows. We also found that for a large momentum $p>5$ GeV, the absolute value of $\lambda_\theta^\text{H}$ is smaller in a stronger magnetic field, which agrees with that shown in Fig. \ref{fig:lambda-theta-B}.

\subsection{Magnetic field perpendicular to momentum}

\begin{figure}
  \centering
  \includegraphics[width=0.6\textwidth]{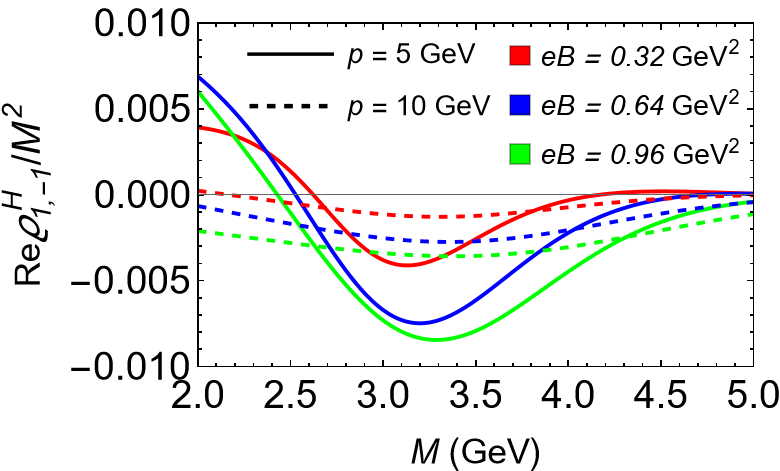}

  \caption{The element $\text{Re}\varrho_{1,-1}^\text{H}$ as a function of invariant mass $M$, at momentum $p=5$ GeV (solid lines) and $p=10$ GeV (dashed lines), calculated in magnetic field with strength $eB=0.32 \text{GeV}^2$ (red lines), $0.64 \text{GeV}^2$ (blue lines), and $0.96 \text{GeV}^2$ (green lines) at temperature $T=0.2$ GeV.}\label{fig:rho-off-spectral}
\end{figure}

We then consider the case that the magnetic field is perpendicular to ${\bf p}$. Without loss of generality, we set the magnetic field along the $z$-direction while  ${\bf p}=(p,0,0)$ is along the $x$-direction. In the helicity frame, the nonvanishing $\lambda$-parameters are $\lambda_\theta^\text{H}$ and $\lambda_\varphi^\text{H}$. The former one is related to the diagonal elements of the spin density matrix, while the latter one is related to $\text{Re}\varrho_{1,-1}$.

For this case, the diagonal elements of the spectral function as a function of invariant mass, magnetic field strength, and magnetic field have similar behavior as the case discussed in the previous subsection, as shown in Fig. \ref{fig:spectral-parallel}. On the other hand, we also have $\text{Re}\,\varrho_{1,-1}^\text{H}\neq 0$ when $eB$ and $p$ are both nonzero values. In Fig. \ref{fig:rho-off-spectral}, we plot $\text{Re}\,\varrho_{1,-1}^\text{H}$ as a function of the invariant mass $M$ for various choices of $eB$ and $p$. In general, $\text{Re}\,\varrho_{1,-1}^\text{H}$ reaches a minimum value at $M\sim3$ GeV - $3.5$ GeV. It preferably to be a negative value except for a small $M$ in a low momentum. Comparing with the diagonal elements of the spectral function shown in Fig. \ref{fig:spectral-parallel}, we find that $\text{Re}\,\varrho_{1,-1}^\text{H}$ is two orders of magnitude smaller than the diagonal elements $\varrho_{\lambda\lambda}$.

\begin{figure}
  \centering
  \includegraphics[width=0.43\textwidth]{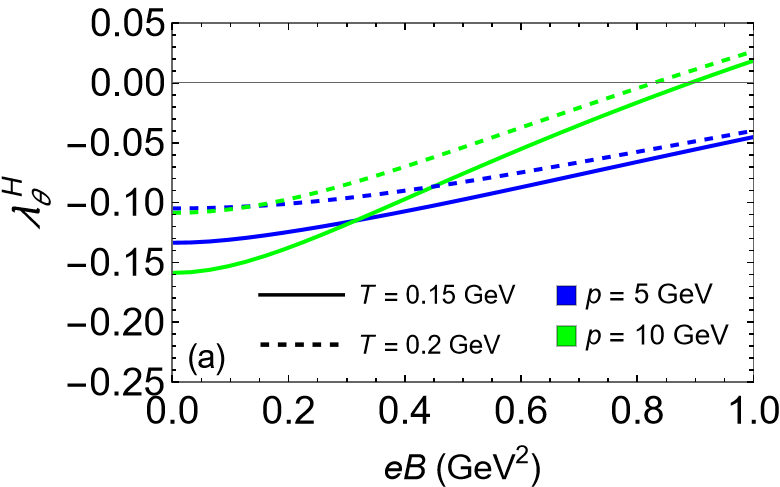}
  \includegraphics[width=0.45\textwidth]{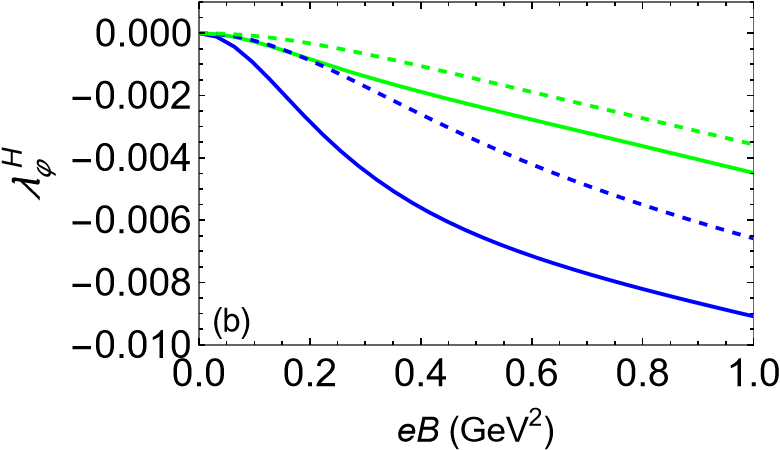}

  \caption{Parameters $\lambda_\theta^\text{H}$ (left panel) and $\lambda_\varphi^\text{H}$ (right panel) as functions of magnetic field strength, at momentum $p=5$ GeV (blue lines) and $p=10$ GeV (green lines), calculated at temperature $T=0.15$ GeV (solid lines) and $0.2$ GeV (dashed lines).}\label{fig:lambda-theta-B-perp}
\end{figure}

In Fig. \ref{fig:lambda-theta-B-perp}, we plot the parameters $\lambda_\theta^\text{H}$ and $\lambda_\varphi^\text{H}$ as functions of magnetic field strength for $J/\psi$ with $p= 5$ GeV and 10 GeV at $T=0.15$ GeV and $T=0.2$ GeV. Comparing Fig. \ref{fig:lambda-theta-B-perp} (a) with Fig. \ref{fig:lambda-theta-B}, we find that the direction of magnetic field does not affect the qualitative behavior for the $B$-dependence of $\lambda_\theta^\text{H}$. On the other hand, the $\lambda_\varphi^\text{H}$ parameter is always zero when $p=0$, and is negative when $p\neq 0$, as shown by  Fig. \ref{fig:lambda-theta-B-perp} (b). This parameter decreases when the field strength increases, but is is one order of magnitude smaller than the $\lambda_\theta^\text{H}$ parameter.

\begin{figure}
  \centering
  \includegraphics[width=0.8\textwidth]{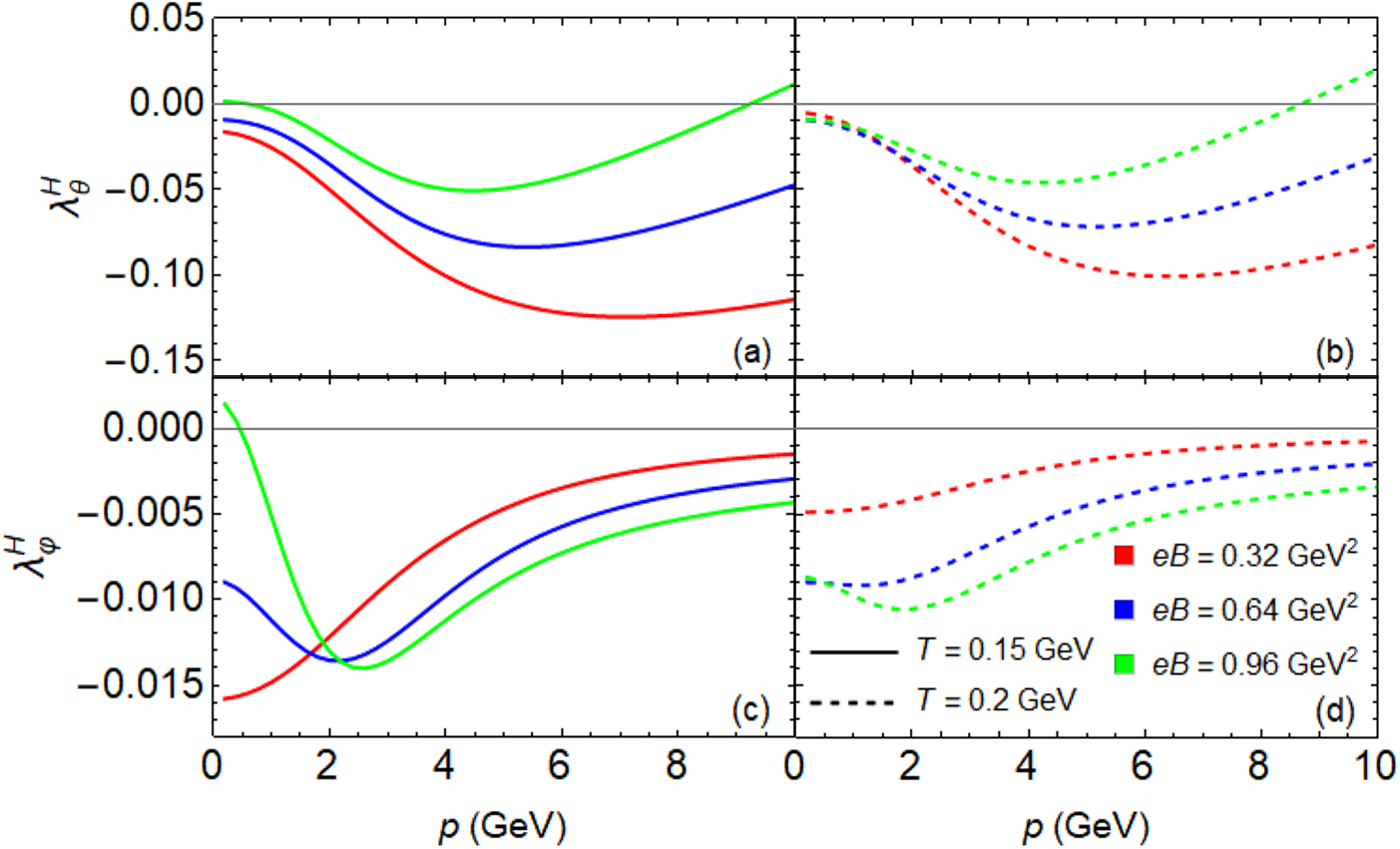}

  \caption{Parameters $\lambda_\theta^\text{H}$ [panels (a) and (b)] and $\lambda_\varphi^\text{H}$ [panels (c) and (d)] as functions of momentum at $eB=0.32 \text{ GeV}^2$ (red lines), $0.64 \text{ GeV}^2$ (blue lines), and $0.96 \text{ GeV}^2$ (green lines), calculated at temperature $T=0.15$ GeV [panels (a) and (c)] and $0.2$ GeV [panels (b) and (d)].}\label{fig:lambda-theta-p-perp}
\end{figure}

The momentum dependence of $\lambda_\theta^\text{H}$ and $\lambda_\varphi^\text{H}$ are shown in Fig. \ref{fig:lambda-theta-p-perp}. We find that in most of the considered regions, both $\lambda_\theta^\text{H}$ and $\lambda_\varphi^\text{H}$ have negative values. The $\lambda_\theta^\text{H}$ parameter first decreases and then increases with increasing $p$. For $0<p<10$ GeV, the absolute value of  $\lambda_\theta^\text{H}$ is smaller in a stronger magnetic field.
On the other hand, the behavior of $\lambda_\varphi^\text{H}$ is more complicated. At $T=0.15$ GeV with $eB=0.32\text{ GeV}^2$, $\lambda_\varphi^\text{H}<0$ when $p\rightarrow0$, and increases towards zero at a larger momentum. However, in stronger magnetic fields, the momentum dependence of $\lambda_\varphi^\text{H}$ is non-monotonic: $\lambda_\varphi^\text{H}$ decreases with increasing $p$, reaches a minimum value at $p\sim2$ GeV, then increases towards zero. The absolute value of $\lambda_\varphi^\text{H}$ at $T=0.2$ is smaller than that at $T=0.15$ GeV, but the momentum dependences are similar. We also find that the $\lambda_\theta^\text{H}$ in Fig. \ref{fig:lambda-theta-p-perp} (a) and (b) is one order of magnitude larger than $\lambda_\varphi^\text{H}$ in Fig. \ref{fig:lambda-theta-p-perp} (c) and (d).

\section{Application to heavy-ion collisions}\label{sec:03}

\begin{figure}
  \centering
  \includegraphics[width=0.8\textwidth]{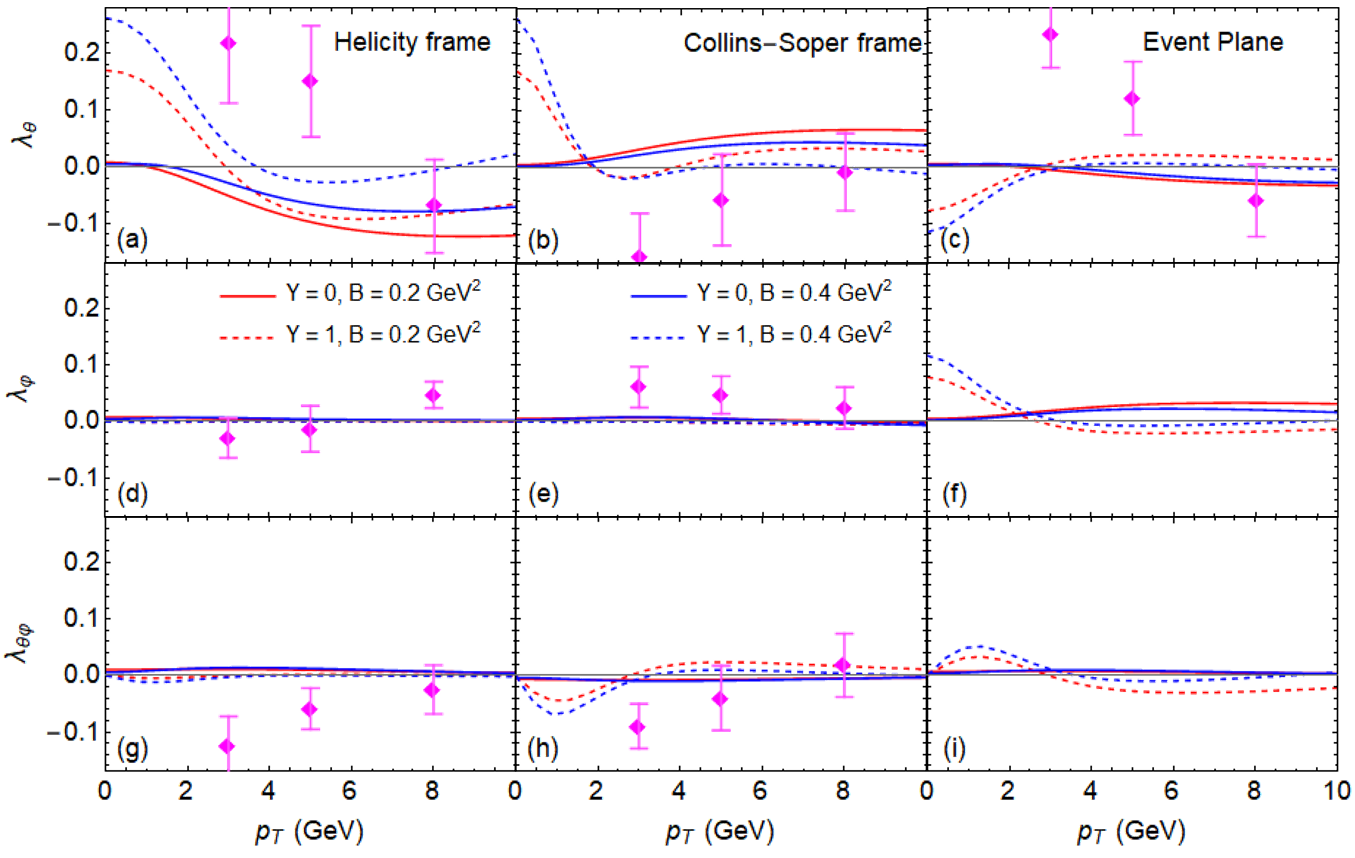}

  \caption{The $\lambda_\theta$, $\lambda_\varphi$, and $\lambda_{\theta\varphi}$ parameters as a function of transverse momentum $p_T$ at $eB=0.2 \text{ GeV}^2$ (red lines) and $eB=0.4 \text{ GeV}^2$ (blue lines), calculated for mesons with rapidity $Y=0$ (solid lines) and $Y=1$ (dashed lines)  at temperature $T=0.15$ GeV.The left, middle, and right columns correspond to the parameters in helicity frame [panels (a), (d), and (g)], Collins-Soper frame [panels (b), (e), and (h)], and those measured along the event-plane direction [panels (c), (f), and (i)], respectively. Magenta diamonds with error bars are experiment results for $J/\psi$ at forward rapidity $2.5<Y<4$ in Pb-Pb collisions at the LHC energy \cite{ALICE:2020iev, ALICE:2022dyy}.}\label{fig:lambda-HIC}
\end{figure}

We now apply our model to real heavy-ion collisions. Considering the magnetic field in noncentral collisions, we fix the direction of magnetic field along the $y$-direction, while the meson's momentum is parameterized by transverse momentum $p_T$, azimuthal angle $\varphi$, and rapidity $Y$,
\begin{equation}
{\bf p}=\left(p_T\cos\varphi,p_T\sin\varphi,\sqrt{M^2+p_T^2}\sinh(Y)\right)
\end{equation}
The analysis is performed for three different spin quantization directions. In the helicity frame, the spin quantization direction is the direction of the vector meson's momentum in the centre-of-mass frame of the collision. In the Collins-Soper frame, the spin quantization direction is defined as the bisector of the angle between the direction of one beam and the opposite direction of the other beam in the rest frame of the vector meson. The third choice is the direction of event plane, i.e., the $y$-direction, in the rest frame. Parameters $\lambda_\theta$, $\lambda_\varphi$, and $\lambda_{\theta\varphi}$ for three cases are shown in Fig. \ref{fig:lambda-HIC} by the first, second, and third columns, respectively. For the helicity frame and the Collins-Soper frame, we find that the $\lambda_\theta$ parameter is dominant while $\lambda_\varphi$ and $\lambda_{\theta\varphi}$ are much smaller than $\lambda_\theta$. But when measuring along the event plane direction, all three parameters $\lambda_\theta^\text{EP}$, $\lambda_\varphi^\text{EP}$, and $\lambda_{\theta\varphi}^\text{EP}$ are of the same order. In Fig. \ref{fig:lambda-HIC}, we have averaged over the meson's azimuthal angle without including the effect of the elliptic flow $v_2$. Using a $v_2$-dependent weight function for the azimuthal angle average only induces negligible corrections to the results.

For mesons at mid rapidity $Y=0$, the $\lambda_\theta$ parameter in the helicity frame and that measured in the event plane direction are negative, while that in the Collins-Soper frame is positive, as shown by solid lines in Fig. \ref{fig:lambda-HIC} (a), (b), and (c). A stronger magnetic field suppresses the absolute value of $\lambda_\theta$ in all three frames. For mesons with $Y=1$, the $\lambda_\theta$ parameter in the helicity frame is positive at low $p_T$, while it decreases and becomes negative at large $p_T$. The  $\lambda_\theta$ parameter in the Collins-Soper frame is also positive for mesons with $Y=1$ and $p_T<2$ GeV, as shown by dashed lines in Fig. \ref{fig:lambda-HIC} (b). It has a negative minimum value at $p_T\sim 3$ GeV. A stronger magnetic field enhances $\lambda_\theta^\text{CS}$ at low $p_T$ but suppresses it at high $p_T$. For the $\lambda_\theta$ in the event plane direction, shown by  Fig. \ref{fig:lambda-HIC} (c), we find that $\lambda_\theta^\text{EP}<0$ for mesons with $Y=1$ and $p_T\apprle 3$ GeV, and  $\lambda_\theta^\text{EP}>0$ when $p_T\apprge 3$ GeV. The $\lambda_\theta^\text{EP}$ parameter decreases in a stronger magnetic field, as indicated by dashed lines in Fig. \ref{fig:lambda-HIC} (c). From Fig. \ref{fig:lambda-HIC} (d) and (e), we find that $\lambda_\varphi$ parameters in the helicity frame and Collins-Soper frame do not have a significant deviation from zero. For the parameters measured in the event-plane direction, we find that $\lambda_\theta\approx-\lambda_\varphi$.

Comparing with the experiment results \cite{ALICE:2020iev, ALICE:2022dyy}, (also shown in Fig. \ref{fig:lambda-HIC} by magenta diamonds with error bars), we find that $\lambda_\theta$ in the helicity frame and Collins-Soper frame, and $\lambda_{\theta\varphi}$ in the Collins-Soper frame for $J/\psi$ with rapidity $Y=1$ qualitatively agree with experiments. The quantitative difference may arise because the experiments were conducted in a more forward rapidity region, $2.5<Y<4$. On the other hand, our model predicts nearly zero $\lambda_\varphi$ parameters in the Helicity frame and the Collins-Soper frame, while the corresponding experiment results also agree with zero within uncertainties. The experiments also observed a significant $\lambda_{\theta\varphi}$ parameter in the helicity frame, which can not be reproduced in our model. For the $\lambda_\theta$ parameter measured along the direction of event plane, we observe a huge gap between our predictions and experiment results, as shown in Fig. \ref{fig:lambda-HIC} (c). This may imply that the theoretical model in this paper could not be directly applied to the forward rapidity region $2.5<Y<4$.

\section{Summary and Outlook}\label{sec:04}
In this article, we employ the framework of gauge/gravity duality to investigate the spin alignment of $J/\psi$ mesons in a thermal-magnetized background. We utilize a soft wall model to describe the QGP background and introduce a massive vector field denoting the probing $J/\psi$ meson.
The meson's spectral function is related to the imaginary part of the two-point current-current correlation, which is derived by solving the equations of motions in the bulk in the gravity theory.

In experiments, the spin of $J/\psi$ is described by five parameters $\lambda_\theta$, $\lambda_\varphi$, $\lambda_{\theta\varphi}$, $\lambda_\varphi^\perp$, and $\lambda_{\theta\varphi}^\perp$, which can be measured by the angular distribution of decay products in the decay process $J/\psi\rightarrow\mu^+\mu^-$. In this paper, we first focus on the case that the meson's momentum is parallel to the magnetic field in the rest frame of the thermal bath. For this case, the meson's spin density matrix is diagonal if we work in the helicity frame, indicating that $\lambda_\theta^\text{H}$ is the only nonvanishing parameter. We analyze through numerical simulations the dependence of spectral functions and $\lambda_\theta^\text{H}$ on the magnetic field strength, meson's momentum, and temperature. We find that the magnetic field induces $\lambda_\theta^\text{H}>0$ when the meson's momentum $p$ is very small, while $\lambda_\theta^\text{H}<0$ when $p$ is large enough. As a comparison, we also study the case of momentum perpendicular to the magnetic field and calculated the spectral functions, $\lambda_\theta^\text{H}$, and $\lambda_\varphi^\text{H}$ as functions of $p$, $eB$, and temperature $T$.

To compare with experiments, we consider the more realistic case with the direction of magnetic field fixed to the $y$-direction while the meson's momentum is determined by the transverse momentum $p_T$, the rapidity $Y$, and the azimuthal angle $\varphi$. For mesons with rapidity $Y=1$, our model prediction shows that $\lambda_\theta$ in the helicity frame decreases with increasing $p_T$, while $\lambda_\theta$ in the Collins-Soper frame is a decreasing (increasing) function of $p_T$ when $p_T<2$ GeV ($p_T>2$ GeV). These results qualitatively agree with experiments \cite{ALICE:2020iev}. Meanwhile, $\lambda_\varphi$ parameters in the helicity and Collins-Soper frames are nearly zero, which also coincide with experiments \cite{ALICE:2020iev} within uncertainties. However, we also find significant differences between our results for $\lambda_{\theta\varphi}^\text{H}$ and $\lambda_\theta^\text{EP}$ with experiments \cite{ALICE:2020iev, ALICE:2022dyy}. The difference may be because of the experiments measure $J/\psi$ at $2.5<Y<4$, while our model studies at $0<Y<1$. As for why we do not consider the same rapidity interval as the experiment: it is because at the centre rapidity region $0<Y<1$, the $J/\psi$ production would be dominated by thermal processes of quark recombinations \cite{Braun-Munzinger:2000csl,ALICE:2022nfv}, which could be effectively described by our model. However, experiment results at $2.5<Y<4$ may be dominated by the initial hard processess\cite{Braun-Munzinger:2000csl,ALICE:2022nfv}, which is beyond the scope of our discussion.

It should be emphasized that the spin alignment of vector meson may exhibit different behavior depending on the explicit type of meson. In our model, the meson's mass plays the role of energy scale and the spin alignment within our model only depends on the parameters, including the normalized magnetic field strength $eB/m^2$ and the normalized temperature $T/m$. For different species of mesons at the same temperature $T$ in a given magnetic field $eB$, these normalized parameters depend on meson's mass and thus the spin alignment may vary according to $m$. As shown in another recent work without magnetic field \cite{Sheng:2024kgg}, $\phi$ and $J/\psi$ mesons could have different preferences of longitudinal and transverse polarizations, respectively, which may qualitatively explain different behaviours of $\phi$ and $J/\psi$ mesons' spin alignments observed by the STAR and ALICE collaborations.

Another interesting point is that the holographic model used in this paper shows the phenomenon of "magnetic catalysis" about chiral condensate. But the phenomenon of ”inverse magnetic catalysis” is observed in lattice QCD simulations\cite{Bali:2012zg}. At first glance, our results may not be convincing. However, whether magnetic field $eB$ acts constructively or destructively on the chiral condensate remains an open question. As recent lattice calculation\cite{Ilgenfritz:2013ara} has shown, the effect of magnetic fields on quark condensate depends on the number of flavors and temperature. Although our holographic model does not explicitly exhibit the number of flavors, within the temperature range studied, the effective mass of the heavy vector meson, which is closely related to quark condensation, shows very weak dependence on the magnetic field(The typical magnetic field strengths generated in the LHC and RHIC experiments are much smaller than the square mass scale of heavy vector mesons ), as shown in Fig.\ref{fig:Mpeak-parallel} and Ref.\cite{Zhao:2023pne}. In addition, our study focuses on the effect of magnetized QGP on the spin alignment of heavy vector mesons(that means it is only related to the nature of the confinement-deconfinement phase transition), rather than discussing the direct influence of the magnetic field on heavy vector mesons. Therefore, even if the magnetic field dependence of chiral condensation varies in different holographic models, this does not significantly affect our main conclusions for the heavy vector mesons.

Based on these intriguing conclusions, we anticipate that there will be more research endeavors in this field in the future. It is well known that in addition to generating a high-temperature and strong magnetic QGP, non-central heavy ion collisions also lead to relatively high QGP density and carry a non-negligible amount of angular momentum. We have investigated the effects of these factors on the current-current correlation function of the $J/\psi$ meson in Ref.\cite{Zhao:2023pne}. Moreover, mesons of different masses exhibit distinct properties, as evident from Ref.\cite{Zhao:2021ogc}, which discusses heavy $J/\psi$ mesons and heavier $\Upsilon(1S)$ mesons. Of course, the properties of lighter mesons are also different from those of heavy mesons, such as $\phi$\cite{Sheng:2022ssp} and $\rho$ mesons. Experimentalists have dedicated substantial time to these matters and are eagerly seeking robust theoretical explanations. Holographic theory, as a potent non-perturbative tool, is expected to provide consistent interpretations for experimental data and reliable theoretical predictions. In addition, considering the fluctuations of all fields and their couplings is an important research direction. In future work, we plan to extend our model to include
these coupling effects by solving the full equations of motion to gain a more comprehensive
understanding of the physical phenomena in magnetized quark-gluon plasma.

\clearpage
\section*{Acknowledgements}
We would like to thank Francesco Becattini and Hai-cang Ren for the encouraging discussions. This work is supported in part by the National Key Research and Development Program of China under Contract No. 2022YFA1604900. This work is also partly supported by the National Natural Science Foundation of China (NSFC) under Grants No. 12275104, and No. 11735007. Si-wen Li is supported by the National Natural Science Foundation of China (NSFC) under Grant No. 12005033 and the Fundamental Research Funds for the Central Universities under Grant No. 3132024192.
	
\appendix
\section{The full action for the vector meson with magnetic field} \label{App1}
Since in the main text, we treat the meson field as a fluctuation neglecting its back-reaction on the background fields (metric and magnetic field), in this appendix, we briefly discuss this setup by taking into account the full action for magnetic field and meson. The theoretical justification for the absence of the direct coupling between the magnetic field and the vector meson field is as follows:

Since vector meson is treated as the non-Abelian part of the generator of flavor group $U\left(N\right)$, we consider two-flavor case i.e. $N=2$ as the most simple example here. In this sense, the total gauge field potential $\mathcal{A}_{M}$ describing both the vector meson field and the magnetic field is the generator of $U\left(2\right)$. Use the decomposition $U\left(2\right)\simeq U\left(1\right)\times SU\left(2\right)$, we have
\begin{equation}
\mathcal{A}_{M}=\hat{A}_{M}+A_{M}.
\end{equation}
Here $\hat{A}_{M}$ is the Abelian generator of $U\left(1\right)$, which is interpreted as the potential of magnetic field, and $A_{M}$ is the non-Abelian generator of $SU\left(2\right)$, which is interpreted as the vector meson field.
Hence the gauge field strength is given by
\begin{align}
\mathcal{F}_{MN}&=\partial_{M}\mathcal{A}_{N}-\partial_{N}\mathcal{A}_{M}+i\left[\mathcal{A}_{M},\mathcal{A}_{N}\right]=\hat{F}_{MN}+W_{MN},
 \end{align}
 where
\begin{align}
  \hat{F}_{MN}&=\partial_{M}\hat{A}_{N}-\partial_{N}\hat{A}_{M},\nonumber\\
   W_{MN}&=F_{MN}+i\left[A_{M},A_{N}\right],\\
   F_{MN}&=\partial_{M}A_{N}-\partial_{N}A_{M}. \nonumber
\end{align}
Altogether, the total action for the gauge field in our case should be
\begin{align}
    S&=-\frac{1}{2}\int d^{4}xd\zeta Q\left(\zeta\right)\mathrm{Tr}\left[\mathcal{F}_{MN}\mathcal{F}^{MN}\right]\nonumber\\
	&=-\frac{1}{2}\int d^{4}xd\zeta Q\left(\zeta\right)\left\{ \frac{1}{2}\hat{F}_{MN}\hat{F}^{MN}+\mathrm{Tr}\left[W_{MN}W^{MN}\right]\right\} \\
	&=-\frac{1}{2}\int d^{4}xd\zeta Q\left(\zeta\right)\left\{ \frac{1}{2}\hat{F}_{MN}\hat{F}^{MN}+\mathrm{Tr}\left[F_{MN}F^{MN}\right]+\mathcal{O}\left(A_{M}^{3}\right)\right\} ,\nonumber
\end{align}
leading to an action including the magnetic part (Abelian) and mesonic part (non-Abelian). Note that $W_{MN}$ is traceless, so $\mathrm{Tr}\left[\hat{F}_{MN}W^{MN}\right]=0$ which is automatically absent in (4), and in our model, the magnetic part
\begin{equation}
    \hat{F}_{MN}\hat{F}^{MN}=\frac{N_{c}}{4\pi^{2}}\left(F_{L}^{2}+F_{R}^{2}\right).
\end{equation}
was considered in Eq. (2.2). Since this part in our model is not much smaller than the metric, we need to solve the background gravity with the back reaction of the magnetic field which is the initial setup of our model. So it is clear that the magnetic field is presented in the total action of the gauge field.
On the other hand, we treat the mesonic part $F_{MN}$ as a probe fluctuation by imposing
\begin{equation}
    g_{MN}>\hat{F}_{MN}>>F_{MN}.
\end{equation}
And we can safely drop the higher order terms $\mathcal{O}(A_M^3)$ because we focus on the two-point holographic correlation function in the current work.

	
\bibliographystyle{utphys}
\bibliography{ref-magnetic}

\end{document}